\documentclass{aa}  

\usepackage{graphicx}
\usepackage{epstopdf}

\usepackage{txfonts}

\begin{document}

   \title{The electron distribution function downstream of the solar-wind
termination shock: Where are the hot electrons?}

\titlerunning{Electrons downstream of the termination shock}
\authorrunning{Fahr et al.}

   \author{Hans J.~Fahr
          \inst{1}
          \and
        John D.~Richardson \inst{2}
        \and
        Daniel Verscharen \inst{3}
          }

   \institute{Argelander Institute for Astronomy, University of Bonn, Auf dem H\"{u}gel 71, 53121 Bonn, Germany\\
              \email{hfahr@astro.uni-bonn.de}
         \and
            Kavli Institute for Astrophysics and Space Research, Massachusetts Institute of Technology, 77 Massachusetts Avenue, Cambridge, MA 02139, USA\\
             \email{jdr@space.mit.edu}
        \and
                Space Science Center, University of New Hampshire, 8 College Road, Durham, NH 03824, USA\\
                \email{daniel.verscharen@unh.edu}
          }

   \date{Received 21 January 2015; accepted 9 May 2015}

 
  \abstract{
In the majority of the literature on plasma shock waves, electrons  play the role of ``ghost particles,'' since their contribution to mass and momentum flows is negligible, and they have been treated as only taking care of the electric plasma neutrality. In some more recent papers, however, electrons play a new important role in the shock dynamics and thermodynamics, especially at  the solar-wind termination shock. They react on the shock electric field in a very specific way, leading to suprathermal nonequilibrium distributions of the downstream electrons, which can be represented by a kappa distribution function. In this paper, we discuss why this anticipated hot electron population has not been seen by the plasma detectors of the Voyager spacecraft  downstream of the solar-wind termination shock. We show that hot nonequilibrium electrons induce a strong negative electric charge-up of any spacecraft cruising through this downstream plasma environment. This charge reduces electron fluxes at the spacecraft detectors to nondetectable intensities. Furthermore, we show that the Debye length $\lambda _{\mathrm D}^{\kappa}$  grows to values of about $\lambda _{\mathrm D}^{\kappa}/\lambda _{\mathrm D}\simeq 10^{6}$ compared to the classical value $\lambda _{\mathrm D}$ in this hot-electron environment. This unusual condition allows for the propagation of a certain type of electrostatic plasma waves that, at very large wavelengths, allow us to determine the effective temperature of the suprathermal electrons directly by means of the phase velocity of these waves. At moderate wavelengths, the electron-acoustic dispersion relation leads to nonpropagating oscillations with the ion-plasma frequency $\omega _{\mathrm p}$ , instead of the traditional electron plasma frequency.}

   \keywords{Plasmas -- Sun: heliosphere -- solar wind -- Shock waves -- Instabilities }

   \maketitle
%

\section{Introduction}

The majority of the plasma-physics literature on shocks essentially considers
general flux-conservation requirements only, leading to the well-known Rankine-Hugoniot relations \citep[e.g., see][]{serrin59,hudson70,baumjohann96,gombosi98,diver01}. These relations, however, do not explicitly formulate the
internal microphysical processes that generate internal entropy during the conversion from the upstream regime into
the downstream plasma. The missing physics in this description
 evidently leads to the well-known phenomenon that the set of
Rankine-Hugoniot relations is a mathematically unclosed system of equations.
Therefore, these relations can only provide unequivocal solutions if
additional physical relations are added to the system, such as  the assumption
of an adiabatic reaction of the plasma ions during their compression into the
higher-density regime on the downstream side \citep[e.g.,][]{erkaev00}.

The solar-wind termination shock is a particular example of a plasma shock for which microphysical effects play an important role. According to recent studies, pick-up ions have a crucial influence on  overall shock physics at the solar-wind termination shock. They are a thermodynamically important additional plasma component, since they extract a significant fraction of the upstream kinetic energy in the form of thermal energy at the termination shock \citep[see][]{decker08}.
\citet{zank10} and \citet{fahr07,fahr10,fahr11a} have studied kinetic features of this multicomponent shock transition and found
relations between upstream and downstream ion distribution functions that are
different for solar-wind protons and pick-up protons. Although these studies discuss the required overadiabatic reaction of pick-up ions,  a satisfying explanation of all plasma properties observed by Voyager-2  \citep{richardson08} is still lacking
as demonstrated by \citet{chalov13}. The latter authors show that assuming a significant difference in the behavior of solar-wind electrons compared to  protons, namely as an independent plasma fluid, leads to an explanation of most of the observed plasma data presented by \citet{richardson08} in a satisfying manner.

To achieve this result in their parameterized study, \citet{chalov13} had to include preferential heating of the solar-wind electrons during the shock passage by a factor of about ten stronger than the proton heating. This type of electron heating at the potential jump of fast-mode shocks had  been realized earlier by \citet{leroy84}, \citet{tokar86}, and \citet{schwartz88}, and the phenomenon of shock-heated electrons also appears in plasma-shock simulations when electrons are treated kinetically \citep[see][]{lembege03,lembege04}. In these cases, the plasma electrons demagnetize due to two-stream and
viscous interactions and attain downstream-to-upstream temperature ratios of 50 and more.

\citet{leroy82} and \citet{goodrich84} follow a different approach. In their treatments, the plasma electrons carry drifts perpendicular to the shock normal ($z$-direction) different from the ion motion as a reaction to the shock-electric field. These drifts establish an electric current $j_{\perp }$ , which is responsible for the change of the surface parallel magnetic field $B_{\parallel }$ in the
form $4\pi j_{\perp }/c=\mathrm dB_{\parallel }/\mathrm dz$. To achieve the same
consistency, \citet{fahr12} describe the conditions of the upstream
and downstream plasma in the bulk frame systems with a frozen-in magnetic field. In this framework, the Liouville-Vlasov theorem describes all relevant downstream plasma quantities as an instantaneous kinetic reaction in the velocity distribution function during the transition from upstream to downstream. The excessive electron heating is then the result of the mass- and charge-specific reactions to the electric shock ramp, as shown in the semikinetic models of the multifluid termination shock by \citet{fahr12} and \citet{fahr13}. According to these studies, electrons enter the downstream side as a strongly heated plasma fluid with negligible mass density that dominates the downstream plasma pressure.

In this paper, we demonstrate why the Voyager-1/-2 spacecraft did not detect these theoretically suggested hot electrons \citep[see][]{richardson08} when they penetrated into the heliosheath plasma. For the purpose of clarification, we analyze the downstream plasma conditions in more detail under which the detection of preferentially heated electrons would have to take place.

\section{Theoretical description of downstream electrons}\label{sect_theor}

In the following section, we shall start from a theoretical description of solar-wind
electrons expected downstream of the termination shock \citep{fahr13}. We treat them as a separate plasma species, which reacts in a very specific manner to the electric-field structure connected with the shock before
adapting to the downstream plasma bulk frame. In the shock-at-rest system, the shock electric potential ramp decelerates the upstream
protons from the upstream bulk velocity $U_{1}$ to the downstream bulk
velocity $U_{2,\mathrm p}$, which is comparable to the center-of-mass flow $U_{2}^{\ast }\simeq U_{2,\mathrm p}(1+s\sqrt{m/M})$, where $s$ is the shock compression ratio and $m,M$ denote the masses of electrons and protons, respectively. The downstream magnetic field is frozen-in into the center-of-mass flow, and all plasma components are eventually comoving with the center-of-mass flow \citep{chashei13}. The electrons, on the other hand, react in a completely different way to this electric potential. First, they attain a strong ``overshoot'' velocity $U_{2,\mathrm e}$ which then relaxes rapidly to
the center-of-mass bulk velocity $U_{2}^{\ast }$ enforced by the frozen-in magnetic
field. During this relaxation process, the plasma generates randomized thermal velocity components  through the action of the two-stream
instability or the Buneman instability as well as by pitch-angle scattering \citep[see][]{chashei13,chashei14,fahr14}.

Under the assumptions of an instantaneous reaction of the electrons to the electric potential and randomization of the overshoot energy by the Buneman instability and pitch-angle scattering to an isotropic distribution in the
downstream bulk frame, we obtain \citep{fahr13,fahr15} the
following expression for the electron pressure $P_{2,\mathrm e}$ on the downstream-side of the shock:
\begin{equation}\label{el_pressure}
P_{2,\mathrm e}=\frac{M}{m}\frac{s^{2}-1}{s}\frac{U_{1}^{2}}{c_{1,\mathrm e}^{2}}\left[A(\alpha
)\sin ^{2}\alpha +B(\alpha )\cos ^{2}\alpha \right]P_{1,\mathrm p}.
\end{equation}
The indices $\mathrm p$ and $\mathrm e$ denote proton- or electron-relevant quantities, and the indices $1$ and $2$ denote upstream and downstream quantities, respectively. The parameter $U$ denotes the
bulk velocity, and $s$ is the shock compression ratio. The parameter $c_{1,\mathrm e}$ is the
average electron thermal velocity on the upstream side, where we assume that solar-wind electrons and protons have equal temperatures. We denote the magnetic tilt angle between the shock normal and the
upstream magnetic field as $\alpha $. The functions $A(\alpha )$ and $B(\alpha )$ are given in \citet{fahr13}. In the limit of vanishing thermal pressures and dominating magnetic pressures, the self-consistent compression ratio $s$ turns out to be $s=1$ \citep[see][Eq.~18]{fahr13}.
Instead of treating the calculated velocity moment $P_{2,\mathrm e}$ of the distribution function, we focus on the distribution function $f_{2,\mathrm e}$ itself  to derive the expected electron particle fluxes $g_{2,\mathrm e}$ as the relevant observable for the Voyager-1/-2 instrumentation. 
As motivated in \citet{fahr13}, we assume that the downstream nonequilibrium distribution function $f_{2,\mathrm e}$ is a general kappa-distribution. This convenient choice represents the transition from a thermal core to a suprathermal tail distribution. The kappa-distribution is given by
\begin{equation}\label{el_distribution}
f_{\mathrm e}(v)=\frac{n_{\mathrm e}}{\pi ^{3/2}\kappa _{\mathrm e}^{3/2}\Theta_{\mathrm e}^3}\frac{\Gamma (\kappa
_{\mathrm e}+1)}{\Gamma (\kappa _{\mathrm e}-1/2)}\left[1+\frac{v^{2}}{\kappa _{\mathrm e}\Theta _{\mathrm e}^{2}}%
\right]^{-\left(\kappa _{\mathrm e}+1\right)},
\end{equation}
where $n_{\mathrm e}$ is the electron number density, $\Theta _{\mathrm e}$ is the velocity
width of the thermal core, and $\kappa _{\mathrm e}$ is the specific electron kappa
parameter with a range of $3/2\leq \kappa _{2,\mathrm e,}\leq \infty$. The symbol 
 $\Gamma =\Gamma (x)$ denotes the Gamma function of the argument 
$x$.

In the next step, we determine the adequate value of $\kappa _{2,\mathrm e}$ for electrons
downstream of the shock associated with a pressure given by Eq.~(\ref{el_pressure}). For
this purpose, we determine the associated pressure $P_{2,\mathrm e,\kappa }$  (i.e., the pressure resulting as the second velocity moment of the above distribution, Eq.~(\ref{el_distribution}), see \citealt{heerikhuisen08}) that is equal to the pressure
given by Eq.~(\ref{el_pressure}) (i.e., the electron pressure found in the multifluid approach by \citet{fahr13}). We obtain the following relation for $P_{2,\mathrm e,\kappa }$:
\begin{multline}\label{el_pressure2}
P_{2,\mathrm e,\kappa }(\kappa )=n_{2,\mathrm e}\frac{m}{2}\Theta _{2,\mathrm e}^{2}\frac{\kappa
_{2,\mathrm e}}{\kappa _{2,\mathrm e}-3/2}\\
=\frac{M}{m}\frac{s^{2}-1}{s}\frac{U_{1}^{2}}{c_{1,\mathrm e}^{2}}[A(\alpha )\sin ^{2}\alpha +B(\alpha )\cos ^{2}\alpha %
]P_{1,\mathrm p}.
\end{multline}
As shown in \citet{fahr13}, we can define the factor $\Pi$ by
\begin{equation}\label{Theta}
\Theta _{2,\mathrm e}^{2}= \Pi \frac{3KT_{1,\mathrm p}}{m}=\Pi \frac{3P_{1,\mathrm p}}{n_{1,\mathrm p}m},
\end{equation}
where $P_{1,\mathrm p}$ is the upstream solar-wind proton pressure. The factor $\Pi $ describes the change of thermal core velocities from upstream to
downstream. We quantify this factor later in Section~\ref{calc_pi}. Using an upstream
proton temperature of $T_{1,\mathrm p}=2\times  10^{4}\,\mathrm K$ and $\Pi\approx 1$, we obtain an average energy for the thermal core electrons of $\left\langle \epsilon _{\mathrm e,\mathrm c}\right\rangle=(1/2)m\Theta _{2,\mathrm e}^{2}=K T_{1,\mathrm p}=1.72\,\mathrm{eV}$. 

From the above relation Eq.~(\ref{el_pressure2}), we first obtain:
\begin{multline}\label{el_density}
\frac{n_{2,\mathrm e}}{2}\Pi \frac{3P_{1,\mathrm p}}{n_{1,\mathrm p}}\frac{\kappa _{2,\mathrm e}}{\kappa
_{2,\mathrm e}-3/2}\\
=\frac{M}{m}\frac{s^{2}-1}{s}\frac{U_{1}^{2}}{c_{1,\mathrm e}^{2}}[A(\alpha )\sin ^{2}\alpha +B(\alpha )\cos ^{2}\alpha ]P_{1,\mathrm p},
\end{multline}
which further simplifies to
\begin{equation}\label{el_density3}
\frac{\kappa _{2,\mathrm e}}{\kappa _{2,\mathrm e}-3/2}=\frac{2}{3}\frac{1}{\Pi }\frac{M}{m}%
\frac{s^{2}-1}{s^{2}}\frac{U_{1}^{2}}{c_{1,\mathrm e}^{2}}\Lambda(\alpha)
\end{equation}
in the region behind the shock, where $U_{2,\mathrm e}\approx U_{2,\mathrm p}$, using the short notation $\Lambda(\alpha)\equiv A(\alpha)\sin^2\alpha +B(\alpha)\cos^2\alpha$.

Relating the thermal velocity $c_{1,\mathrm e}$ of the upstream electrons to the thermal velocity $c_{1,\mathrm p}$ of the upstream
protons through $T_{1,\mathrm p}=T_{1,\mathrm e}$, we obtain
\begin{equation}\label{el_density4}
\frac{\kappa _{2,\mathrm e}}{\kappa _{2,\mathrm e}-3/2}=\frac{2}{3\Pi }\frac{s^{2}-1}{s^{2}}%
\frac{U_{1}^{2}}{c_{1,p}^{2}}\Lambda(\alpha).
\end{equation}
We assume that the upstream solar-wind Mach number of the protons (i.e., $\mu _{1,\mathrm p}\equiv U_{1}/c_{1,\mathrm p}$) is of  order 8, which then leads to
\begin{equation}\label{el_kappa}
\kappa _{2,\mathrm e}=\frac{3}{2}\frac{\frac{1}{\Pi }\frac{s^{2}-1}{s^{2}}%
8^{2}\Lambda (\alpha )}{\frac{1}{\Pi }\frac{s^{2}-1}{s^{2}}8^{2}\Lambda
(\alpha )-3/2}.
\end{equation}
The termination-shock compression ratio observed by Voyager-2 is $s\simeq 2.5$ \citep{richardson08}, which leads to
\begin{equation}\label{el_kappa2}
\kappa _{2,\mathrm e}=\frac{3}{2}\frac{54\Lambda (\alpha )/\Pi }{54\Lambda (\alpha
)/\Pi -3/2}.
\end{equation}
In the case of a perpendicular shock (i.e., $\alpha \simeq 90^{\circ }$), $\Lambda (\pi /2)=A(\pi /2)=s$ leading to
\begin{equation}\label{el_kappa3}
\kappa _{2,\mathrm e}=\frac{3}{2}\frac{135/\Pi }{135/\Pi -3/2}.
\end{equation}
Assuming a value of $\Pi =1$, we obtain the result $\kappa _{2,\mathrm e}=1.517$. This kappa
index $\kappa _{2,\mathrm e}$ for the shocked downstream solar-wind electrons
characterizes a highly suprathermal electron spectrum with a power-law
nearly falling off as $v^{-5}$ as shown in Eq.~(\ref{el_distribution}). Consequently, we can write for the resulting distribution function of the downstream electrons,
\begin{equation}\label{el_distribution3}
f_{2,\mathrm e}(v)=\frac{n_{2,\mathrm e}}{\pi ^{3/2}\kappa _{2,\mathrm e}^{3/2}\Theta_{2,\mathrm e}^3}\frac{\Gamma
(\kappa _{2,\mathrm e}+1)}{\Gamma (\kappa _{2,\mathrm e}-1/2)}\left[1+\frac{v^{2}}{\kappa_{2,\mathrm e}\Theta _{2,\mathrm e}^{2}}\right]^{-(\kappa _{2,\mathrm e}+1)}
\end{equation}
with $\kappa _{2,\mathrm e}\simeq 1.517$.

\begin{figure}
 \resizebox{\hsize}{!}{\includegraphics{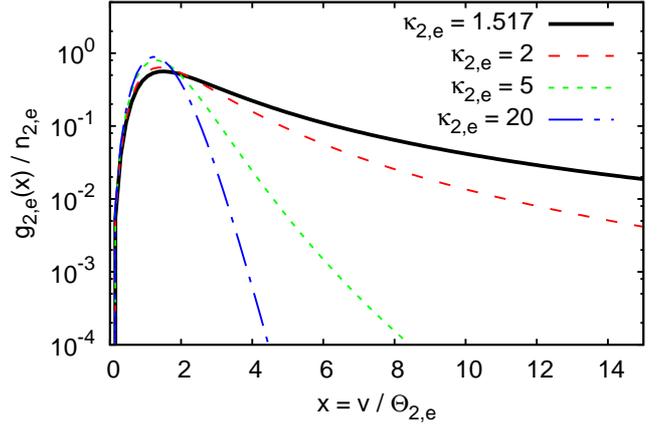}}
  \caption{Normalized spectral electron fluxes downstream of the solar-wind termination shock according to Eq.~(\ref{g2e}). We show the kappa-distributed fluxes for different values of $\kappa_{2,\mathrm e}$. The special case $\kappa_{2,\mathrm e}=1.517$ is our result for the perpendicular solar-wind termination shock according to Eq.~(\ref{el_kappa3}).}
  \label{figure1}
\end{figure}

The distribution function in Eq.~(\ref{el_distribution3}) is easily transformed into the spectral electron flux $g_{2,\mathrm e}(v)=4\pi v^{3}f_{2,\mathrm e}(v)$. We normalize velocities as $x=v/\Theta _{2,\mathrm e}$ and find
\begin{equation}\label{g2e}
g_{2,\mathrm e}(x)=\frac{4n_{2,\mathrm e}}{\pi ^{1/2}\kappa _{2,\mathrm e}^{3/2}}\frac{\Gamma
(\kappa _{2,\mathrm e}+1)}{\Gamma (\kappa _{2,\mathrm e}-1/2)}x^{3}\left[1+\frac{x^{2}}{\kappa
_{2,\mathrm e}}\right]^{-(\kappa _{2,\mathrm e}+1)}.
\end{equation}
In Figure~\ref{figure1}, we show these spectral electron fluxes for
different indices $\kappa _{2,\mathrm e}$.

\section{Electric equilibrium potential}

In the following, we calculate the electric equilibrium potential $\Phi $ up to which a spacecraft charges up when entering the heliosheath under the assumption that both electrons and protons have kappa-distributions and that emission processes are negligible in the plasma. \citet{fahr13} show that all ions (solar-wind as well as pick-up ions) treated as one fluid can be characterized as one joint kappa-function with a joint kappa-index $\kappa _{2,\mathrm i}\simeq 2,$ depending on the pick-up ion abundance downstream of the shock \citep[see Figure~2 of ][]{fahr13}. Therefore, we describe  electrons with Eq.~(\ref{el_distribution3}) and protons with the following distribution downstream of the shock:
\begin{equation}
f_{2,\mathrm i}(v)=\frac{n_{2,\mathrm i}}{\pi ^{3/2}\kappa _{2,\mathrm i}^{3/2}\Theta_{2,\mathrm i}^3}\frac{\Gamma
(\kappa _{2,\mathrm i}+1)}{\Gamma (\kappa _{2,\mathrm i}-1/2)}\left[1+\frac{v^{2}}{\kappa
_{2,\mathrm i}\Theta _{2,\mathrm i}^{2}}\right]^{-(\kappa _{2,\mathrm i}+1)}.
\end{equation}
We assume that  quasineutrality prevails outside of perturbed Debye regions, i.e., $n_{2,\mathrm e}=n_{2,\mathrm i}$.

Any metallic body embodied in the heliosheath plasma downstream of the shock charges up to an electric equilibrium
potential $\Phi _{2}$ , which  guarantees equal fluxes of ions and electrons reaching the metallic surface of this body per unit time
(geometry taken to be planar). This behavior leads to the following requirement (after dropping downstream indices ``2'' for simplification):
\begin{multline}
\beta _{\mathrm e}(\Phi )\iiint (v\cos \theta
)f_{\mathrm e}(v)v^{2}\,\mathrm dv\,\mathrm d\phi \sin \theta\,\mathrm d\theta \\
=\beta _{\mathrm i}(\Phi )\iiint (v\cos \theta )f_{\mathrm i}(v)v^{2}\,\mathrm dv\,\mathrm d\phi \sin \theta\,\mathrm d\theta,
\end{multline}
where $\beta _{\mathrm e}(\Phi )$ and $\beta _{\mathrm i}(\Phi )$ denote the Boltzmann
screening factors for electrons and ions, respectively. These factors describe the fraction of particles that can reach the wall against the electric potential $\Phi$. The assumption of isotropic distribution functions leads to
\begin{equation}
\beta _{\mathrm e}(\Phi )\int f_{\mathrm e}(v)v^{3}\,\mathrm dv=\beta _{\mathrm i}(\Phi )\int
f_{\mathrm i}(v)v^{3}\,\mathrm dv.
\end{equation}
We expect that the resulting equilibrium potential $\Phi$ only affects the lowest-energy part of the distribution functions. Therefore, the  Gaussian core of the kappa-distributed particles is the only screened population, leading to
\begin{equation}
\frac{\beta _{\mathrm e}(\Phi )}{\beta _{\mathrm i}(\Phi )}=\frac{\exp \left(+2e\Phi /m\Theta
_{\mathrm e}^{2}\right)}{\exp \left(-2e\Phi /M\Theta _{\mathrm i}^{2}\right)}=\frac{\int f_{\mathrm i}(v)v^{3}\,\mathrm dv}{\int f_{\mathrm e}(v)v^{3}\,\mathrm dv},
\end{equation}
which can be rewritten as
\begin{equation}
\exp \left[2e\Phi \left(\frac{1}{m\Theta _{\mathrm e}^{2}}+\frac{1}{M\Theta _{\mathrm i}^{2}}\right)\right]=
\frac{\int f_{\mathrm i}(v)v^{3}\,\mathrm dv}{\int f_{\mathrm e}(v)v^{3}\,\mathrm dv}.
\end{equation}
We solve the remaining integrals in the above expression with $\Theta_{2,\mathrm i}/\Theta_{2,\mathrm e}\approx m/M$, leading to
\begin{equation}
\frac{\int f_{\mathrm i}(v)v^{3}\,\mathrm dv}{\int f_{\mathrm e}(v)v^{3}\,\mathrm dv}=\frac{\Gamma (\kappa
_{\mathrm i}+1)}{\Gamma (\kappa _{\mathrm i}-1/2)}\frac{\Gamma (\kappa _{\mathrm e}-1/2)}{\Gamma
(\kappa _{\mathrm e}+1)}\frac{\sqrt{\kappa_{\mathrm e}m}\left(\kappa_{\mathrm e}-1\right)}{\sqrt{\kappa_{\mathrm i}M}\left(\kappa_{\mathrm i}-1\right)}.
\end{equation}
We find (with $\Pi=\Theta_{2,\mathrm e}^2/\Theta_{1,\mathrm e}^2\approx 1$)
\begin{equation}
\exp \left[\frac{2e\Phi }{mU_{1}^{2}}\left(\frac{U_{1}^{2}}{\Theta _{\mathrm e}^{2}}+\frac{%
mU_{1}^{2}}{M\Theta _{\mathrm i}^{2}}\right)\right]=\exp \left[\frac{2e\Phi }{MU_{1}^{2}}\left(\mu
_{1,\mathrm e}^{2}+\mu _{1,\mathrm i}^{\ast 2}\right)\right],
\end{equation}
where $\mu _{1,\mathrm e}$ and $\mu _{1,\mathrm i}^{\ast }$ denote the upstream solar-wind
electron and pick-up ion Mach numbers. These numbers are given by values
of the order $\mu _{1,\mathrm e}\simeq 10$ and $\mu _{1,\mathrm i}^{\ast }\simeq 3$. Therefore,
\begin{equation}
\exp \left(\frac{2e\Phi }{MU_{1}^{2}}10^{2}\right)=\frac{\Gamma (\kappa
_{\mathrm i}+1)}{\Gamma (\kappa _{\mathrm i}-1/2)}\frac{\Gamma (\kappa _{\mathrm e}-1/2)}{\Gamma
(\kappa _{\mathrm e}+1)}\frac{\sqrt{\kappa_{\mathrm e}m}\left(\kappa_{\mathrm e}-1\right)}{\sqrt{\kappa_{\mathrm i}M}\left(\kappa_{\mathrm i}-1\right)},
\end{equation}
which leads to the following potential
\begin{equation}\label{phi}
\Phi =\frac{MU_{1}^{2}}{200e}\ln \left[\frac{\Gamma (\kappa
_{\mathrm i}+1)}{\Gamma (\kappa _{\mathrm i}-1/2)}\frac{\Gamma (\kappa _{\mathrm e}-1/2)}{\Gamma
(\kappa _{\mathrm e}+1)}\frac{\sqrt{\kappa_{\mathrm e}m}\left(\kappa_{\mathrm e}-1\right)}{\sqrt{\kappa_{\mathrm i}M}\left(\kappa_{\mathrm i}-1\right)} \right].
\end{equation}
As a consistency check, we note that this expression leads to the classical plasma-physics formula for $\Phi =\Phi _{\mathrm c}$ in the limit of Maxwellian distributions (i.e., $\kappa _{\mathrm e}=\kappa_{\mathrm i}\rightarrow \infty $) with identical temperatures $T_{\mathrm i}=T_{\mathrm e}=MU_{1}^{2}/200K$.

\begin{figure}
 \resizebox{\hsize}{!}{\includegraphics{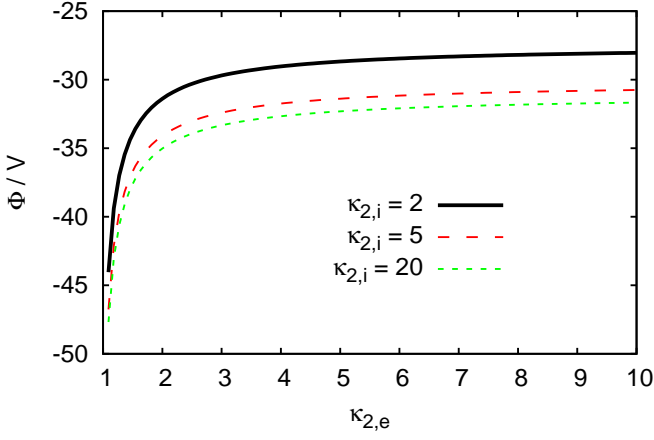}}
  \caption{Equilibrium potential of a metallic body in the presence of kappa-distributed ions and electrons according to Eq.~(\ref{phi}). We show the potential as a function of $\kappa_{2,\mathrm e}$ for different values of $\kappa_{2,\mathrm i}$. We assume $U_1=400\,\mathrm{km/s}$.}
  \label{figure2}
\end{figure}

In Figure~\ref{figure2}, we show the resulting electric potential $\Phi $ as a function of the prevailing kappa-index $\kappa _{2,\mathrm e}$ of the shock-heated downstream electrons. This profile shows that the expected equilibrium potential
drops to values of $\Phi \leq -30\,\mathrm V$ in the range of expected indices $1.5\leq \kappa _{\mathrm e}\leq 2$. This potential
does not allow electrons with energies below $30\,\mathrm{eV}$ to reach the detector.

\section{Degenerated Debye length}

The degeneration of the Debye length is a direct consequence of highly nonthermal kappa-type distribution
functions. The electric screening by plasma-electron distributions with a largely extended power-law tail is significantly less efficient compared to the screening by Maxwellian thermal electrons with temperature $T_{\mathrm e}$. Maxwellian thermal electrons lead to the classical Debye-screening length of $\lambda _{\mathrm D}=\sqrt{KT_{\mathrm e}/4\pi ne^{2}}$.
The effect of degenerating Debye lengths has been recognized and emphasized by \citet{treumann04}, finding that  the resulting Debye length $\lambda_{\mathrm D}^{\kappa }$ may easily increase by factors of $\lambda _{\mathrm D}^{\kappa
}/\lambda _{\mathrm D}\geq 10^{6}$ for low electron kappa-indices of $\kappa _{\mathrm e}\simeq 1.5$. With a more relaxed approach, yet along the lines of these authors' discussion, we obtain the following very similar conclusions.

For the description of the effective screening of kappa electrons, we simply replace the Maxwellian temperature $T_{\mathrm e}$ by the corresponding electron kappa temperature $T_{\mathrm e}^{\kappa }$, which is given by
\begin{equation}\label{kTkappa}
KT_{\mathrm e}^{\kappa }=\frac{P_{\mathrm e}^{\kappa }}{n_{\mathrm e}}=\frac{m}{2}\Theta_{\mathrm e} ^{2}\frac{%
\kappa _{\mathrm e}}{\kappa _{\mathrm e}-3/2}.
\end{equation}
Consequently,  we find
\begin{equation}
\lambda _{\mathrm D}^{\kappa }/\lambda _{\mathrm D}=\sqrt{T_{\mathrm e}^{\kappa }/T_{\mathrm e}}=\sqrt{\frac{m}{2}\Theta_{\mathrm e} ^{2}\frac{\kappa _{\mathrm e}}{\kappa _{\mathrm e}-3/2}/KT_{\mathrm e}}.
\end{equation}
Assuming that the core of the kappa distribution is identical to the
Maxwellian core (i.e., $m\Theta_{\mathrm e} ^{2}=2KT_{\mathrm e}$), we
obtain  the following result for the effective Debye length:
\begin{equation}\label{lambdakappa}
\lambda _{\mathrm D}^{\kappa }=\lambda _{\mathrm D}\sqrt{\frac{\kappa _{\mathrm e}}{\kappa _{\mathrm e}-3/2}}.
\end{equation}

This modified Debye length has an interesting effect on the propagation of plasma waves. The general dispersion relation for
electron-acoustic plasma waves \citep[e.g., see][Eq.~(4-48)]{chen74} is given by
\begin{equation}
\frac{\omega }{k}=\sqrt{\frac{KT_{\mathrm e}^{\kappa }}{M}\frac{1}{1+\left(k\lambda
_{\mathrm D}^{\kappa }\right)^{2}}+\frac{KT_{\mathrm i}}{M}}.
\end{equation}
In the general case of $T_{\mathrm e}^{\kappa }\gg T_{i}$ , and for very
small wavevector values of $k\equiv 2\pi /\lambda \ll 2\pi /\lambda _{\mathrm D}^{\kappa }$,
this dispersion relation allows for a branch of electron-acoustic waves that propagate with a phase/group velocity of
\begin{equation}
\frac{\omega }{k}=\frac{\partial \omega }{\partial k}=\sqrt{\frac{%
KT_{\mathrm e}^{\kappa }}{M}}=\sqrt{\frac{m}{2M}\Theta _{\mathrm e}^{2}\frac{\kappa _{\mathrm e}}{\kappa
_{\mathrm e}-3/2}}.
\end{equation}
This branch is a special kappa-mode propagating with a typical phase or group velocity that directly depends on the electron kappa index $\kappa _{\mathrm e}$. Testing plasma acoustic waves in this range of large wavelengths should hence directly
reveal the prevailing kappa index $\kappa _{\mathrm e}$ and, therefore, the character of the suprathermal downstream electrons.

On the other hand, the limit of larger wavevectors with $k\equiv 2\pi /\lambda \geq 2\pi
/\lambda _{\mathrm D}^{\kappa }$ allows for a branch of nonpropagating waves (i.e., standing oscillations) with plasma eigenfrequencies given by
\begin{equation}
\frac{\omega }{k}=\sqrt{\frac{KT_{\mathrm e}^{\kappa }}{M}\frac{1}{\left(k\lambda
_{\mathrm D}^{\kappa }\right)^{2}}}.
\end{equation}
With Eqs.~(\ref{kTkappa}) and (\ref{lambdakappa}), we can write 
\begin{equation}
\omega =\sqrt{\frac{KT_{\mathrm e}^{\kappa }}{M\lambda _{\mathrm D}^{\kappa 2}}}=\sqrt{%
\frac{\frac{m}{2}\Theta _{\mathrm e}^{2}\frac{\kappa _{\mathrm e}}{\kappa _{\mathrm e}-3/2}}{M\lambda
_{\mathrm D}^{2}\left(\frac{\kappa _{\mathrm e}}{\kappa _{\mathrm e}-3/2}\right)}}=\sqrt{\frac{4\pi ne^{2}}{M}}%
=\omega _{\mathrm p}
\end{equation}
for these oscillations. Under these conditions, the plasma, surprisingly enough, does not oscillate with the electron plasma frequency $\omega _{\mathrm e}$, but it does oscillate with the ion plasma frequency $\omega _{\mathrm p}$, which is a phenomenon that only quite rarely
occurs in nature. The plasma oscillations recently registered by Voyager-1 \citep{gurnett13}, which
were interpreted as electron plasma oscillations, may perhaps be reinterpreted as this type of ion plasma oscillations. As such, they 
would allow us to infer environmental plasma densities of the order $n\simeq (m/M)0.1\,\mathrm{cm}^{-3}\leq 10^{-4}\,\mathrm{cm}^{-3}$.

\section{Calculation of $\Pi$ in view of the downstream electron instabilities}\label{calc_pi}

In Section~\ref{sect_theor}, we introduced the quantity $\Pi $ (see Eq.~(\ref{Theta})), which
denotes the ratio of the thermal core widths, $\Pi =\Theta _{2,\mathrm e}^{2}/\Theta _{1,\mathrm e}^{2}$. Until this point, we
did not determine a reasonable value for $\Pi$.
Previous expressions based on an upstream-downstream transformation of thermal-core velocities \citep{fahr13} appear to be irrelevant since their calculation relies on the assumption that upstream core electrons are independently transformed simply into downstream core electrons according to the Liouville-Vlasov theorem. In reality, however, all upstream electrons overshoot to the downstream side where the action of instabilities, such as the two-stream instability or the Buneman instability, redistribute and isotropize them. In the case of the two-stream instability \citep[e.g., see][]{chen74}, electrons can excite ion
oscillations as long as their velocities are greater than the thermal core velocities of the protons. This fast relaxation of the electron distribution function toward the downstream ion distribution function  then leads to a quasiequilibrium distribution.

Consistent forms of such quasi-equilibria between particle distribution
functions and turbulence power spectra have been investigated in Section 2.5 in \citet{fahr11b} and, for steady state conditions, by \citet{yoon11,yoon12} and \citet{zaheer13}. In all of these cases, the asymptotic state results in kappa distributions. Also, in our case, as a result of a shock-induced electron injection with velocity-space diffusion and relaxation described by a phase-space transport
equation of the type
\begin{equation}
\frac{\partial f_{\mathrm e}}{\partial t}=\frac{1}{v^{2}}\frac{\partial }{\partial v}%
\left(v^{2}D_{vv}\frac{\partial f_{\mathrm e}}{\partial v}\right)+\frac{f_{\mathrm e}-f_{\mathrm p}}{\tau _{\mathrm{ep}}},
\end{equation}
we expect solutions in form of kappa distributions. In fact, as shown by \citet{treumann04}, this kind of  transport equations leads to a
quasi-equilibrium distribution in the form of a kappa distribution with a
thermal core given by the downstream ion velocities: $\Theta
_{2,\mathrm e}^{2}(f_{\mathrm p})\simeq 2P_{2,\mathrm p}/n_{2,\mathrm e}m$. We finally find  \citep[using Eq.~(4) in][]{fahr13}
\begin{equation}
\Pi =\Theta _{2,\mathrm e}^{2}/\Theta _{1,\mathrm e}^{2}=\frac{2P_{2,\mathrm p}n_{1}m}{%
3P_{1,\mathrm p}n_{2,\mathrm e}m}=\frac{2}{9}\left[2A(s,\alpha )+B(s,\alpha )\right]
\end{equation}
with $A(s,\alpha )=\sqrt{\cos ^{2}\alpha +s^{2}\sin ^{2}\alpha }$ and $B(s,\alpha )=s^{2}/A^{2}(s,\alpha )$.

The above expression for a perpendicular shock with $s=2.5$ \citep{richardson08} leads to $\Pi (\alpha =\pi /2)=1.33$.
This finally shows that the quantity $\Pi$ is, in fact, of order unity, verifying \emph{\textup{a posteriori}} all of our above results that we calculated for $\Pi =1$. Using this more precise value for $\Pi$, Eq.~(\ref{el_kappa3})  leads to a marginally different value for the kappa-index of $\kappa _{2,\mathrm e}=1.522$.

\section{Summary and conclusions}

Previous studies suggest that strongly-heated solar-wind electrons should
appear in measurements as accelerated suprathermal particles downstream of the termination
shock. However, these electrons were not observed by Voyager in the
heliosheath. We investigate this apparent contradiction and find that
heliosheath electrons are distributed according to a kappa-type distribution function with an extended suprathermal tail (see
Figure 1). These highly suprathermal kappa-distributed electrons lead to a strong negative charging of all metallic bodies exposed to this plasma environment, consequently also charging up the Voyager spacecraft.

A spacecraft potential of the order $-30\,\mathrm V,$ as calculated in Figure~\ref{figure2}, has a significant effect on the Voyager electron  measurements in the heliosheath. Under these conditions, it repels thermal electrons in the
energy range below 30 eV, leading to an increase in the previously determined upper limit of 3 eV
\citep{richardson08} for the electron temperature. The ions, on the other hand, are accelerated into the Faraday cups. The difference in the derived bulk speeds, however, is negligible: corrected for a -30~V potential, the radial downstream proton bulk velocity increases from 130 km/s to about 132 km/s.

In addition, this suprathermal distribution of downstream electrons also
results in an unusually enlarged Debye length. As a consequence of this effect, the phase velocity $v_{\phi
}=\omega /k$ of electrostatic plasma waves depends on the effective kappa temperature of the electrons in the heliosheath plasma environment. The detection of these plasma waves allows us to infer the effective kappa electron temperature as an observable quantity. These distributions also permit a type of nonpropagating standing waves with the ion plasma frequency $\omega _{\mathrm p}$ as their eigen frequency.

\bibliographystyle{aa}
\bibliography{electrons_shocked}

\begin{thebibliography}{34}
\expandafter\ifx\csname natexlab\endcsname\relax\def\natexlab#1{#1}\fi

\bibitem[{{Baumjohann} \& {Treumann}(1996)}]{baumjohann96}
{Baumjohann}, W. \& {Treumann}, R.~A. 1996, {Basic space plasma physics}

\bibitem[{{Chalov} \& {Fahr}(2013)}]{chalov13}
{Chalov}, S.~V. \& {Fahr}, H.~J. 2013, \mnras, 433, L40

\bibitem[{{Chashei} \& {Fahr}(2013)}]{chashei13}
{Chashei}, I.~V. \& {Fahr}, H.~J. 2013, Annales Geophysicae, 31, 1205

\bibitem[{{Chashei} \& {Fahr}(2014)}]{chashei14}
{Chashei}, I.~V. \& {Fahr}, H.~J. 2014, \solphys, 289, 1359

\bibitem[{{Chen}(1974)}]{chen74}
{Chen}, F.~F. 1974, {Introduction to plasma physics}

\bibitem[{{Decker} {et~al.}(2008){Decker}, {Krimigis}, {Roelof}, {Hill},
  {Armstrong}, {Gloeckler}, {Hamilton}, \& {Lanzerotti}}]{decker08}
{Decker}, R.~B., {Krimigis}, S.~M., {Roelof}, E.~C., {et~al.} 2008, \nat, 454,
  67

\bibitem[{{Diver}(2001)}]{diver01}
{Diver}, D.~A. 2001, {A plasma formulary for physics, technology, and
  astrophysics}

\bibitem[{{Erkaev} {et~al.}(2000){Erkaev}, {Vogl}, \& {Biernat}}]{erkaev00}
{Erkaev}, N.~V., {Vogl}, D.~F., \& {Biernat}, H.~K. 2000, Journal of Plasma
  Physics, 64, 561

\bibitem[{{Fahr} {et~al.}(2014){Fahr}, {Chashei}, \& {Verscharen}}]{fahr14}
{Fahr}, H.~J., {Chashei}, I.~V., \& {Verscharen}, D. 2014, \aap, 571, A78

\bibitem[{{Fahr} \& {Fichtner}(2011)}]{fahr11b}
{Fahr}, H.~J. \& {Fichtner}, H. 2011, \aap, 533, A92

\bibitem[{{Fahr} \& {Siewert}(2007)}]{fahr07}
{Fahr}, H.-J. \& {Siewert}, M. 2007, Astrophysics and Space Sciences
  Transactions, 3, 21

\bibitem[{{Fahr} \& {Siewert}(2010)}]{fahr10}
{Fahr}, H.-J. \& {Siewert}, M. 2010, \aap, 512, A64

\bibitem[{{Fahr} \& {Siewert}(2011)}]{fahr11a}
{Fahr}, H.-J. \& {Siewert}, M. 2011, \aap, 527, A125

\bibitem[{{Fahr} \& {Siewert}(2013)}]{fahr13}
{Fahr}, H.-J. \& {Siewert}, M. 2013, \aap, 558, A41

\bibitem[{{Fahr} \& {Siewert}(2015)}]{fahr15}
{Fahr}, H.-J. \& {Siewert}, M. 2015, \aap, 576, A100

\bibitem[{{Fahr} {et~al.}(2012){Fahr}, {Siewert}, \& {Chashei}}]{fahr12}
{Fahr}, H.-J., {Siewert}, M., \& {Chashei}, I. 2012, \apss, 341, 265

\bibitem[{{Gombosi}(1998)}]{gombosi98}
{Gombosi}, T.~I., ed. 1998, {Physics of the space environment}

\bibitem[{{Goodrich} \& {Scudder}(1984)}]{goodrich84}
{Goodrich}, C.~C. \& {Scudder}, J.~D. 1984, \jgr, 89, 6654

\bibitem[{{Gurnett} {et~al.}(2013){Gurnett}, {Kurth}, {Burlaga}, \&
  {Ness}}]{gurnett13}
{Gurnett}, D.~A., {Kurth}, W.~S., {Burlaga}, L.~F., \& {Ness}, N.~F. 2013, AGU
  Fall Meeting Abstracts, B1

\bibitem[{{Heerikhuisen} {et~al.}(2008){Heerikhuisen}, {Pogorelov},
  {Florinski}, {Zank}, \& {le Roux}}]{heerikhuisen08}
{Heerikhuisen}, J., {Pogorelov}, N.~V., {Florinski}, V., {Zank}, G.~P., \& {le
  Roux}, J.~A. 2008, \apj, 682, 679

\bibitem[{{Hudson}(1970)}]{hudson70}
{Hudson}, P.~D. 1970, \planss, 18, 1611

\bibitem[{{Lemb{\`e}ge} {et~al.}(2004){Lemb{\`e}ge}, {Giacalone}, {Scholer},
  {Hada}, {Hoshino}, {Krasnoselskikh}, {Kucharek}, {Savoini}, \&
  {Terasawa}}]{lembege04}
{Lemb{\`e}ge}, B., {Giacalone}, J., {Scholer}, M., {et~al.} 2004, \ssr, 110,
  161

\bibitem[{{Lemb{\`e}ge} {et~al.}(2003){Lemb{\`e}ge}, {Savoini}, {Balikhin},
  {Walker}, \& {Krasnoselskikh}}]{lembege03}
{Lemb{\`e}ge}, B., {Savoini}, P., {Balikhin}, M., {Walker}, S., \&
  {Krasnoselskikh}, V. 2003, \jgr, 108, 1256

\bibitem[{{Leroy} \& {Mangeney}(1984)}]{leroy84}
{Leroy}, M.~M. \& {Mangeney}, A. 1984, Annales Geophysicae, 2, 449

\bibitem[{{Leroy} {et~al.}(1982){Leroy}, {Winske}, {Goodrich}, {Wu}, \&
  {Papadopoulos}}]{leroy82}
{Leroy}, M.~M., {Winske}, D., {Goodrich}, C.~C., {Wu}, C.~S., \&
  {Papadopoulos}, K. 1982, \jgr, 87, 5081

\bibitem[{{Richardson} {et~al.}(2008){Richardson}, {Kasper}, {Wang}, {Belcher},
  \& {Lazarus}}]{richardson08}
{Richardson}, J.~D., {Kasper}, J.~C., {Wang}, C., {Belcher}, J.~W., \&
  {Lazarus}, A.~J. 2008, \nat, 454, 63

\bibitem[{{Schwartz} {et~al.}(1988){Schwartz}, {Thomsen}, {Bame}, \&
  {Stansberry}}]{schwartz88}
{Schwartz}, S.~J., {Thomsen}, M.~F., {Bame}, S.~J., \& {Stansberry}, J. 1988,
  \jgr, 93, 12923

\bibitem[{{Serrin}(1959)}]{serrin59}
{Serrin}, J. 1959, Handbuch der Physik, 8, 125

\bibitem[{{Tokar} {et~al.}(1986){Tokar}, {Aldrich}, {Forslund}, \&
  {Quest}}]{tokar86}
{Tokar}, R.~L., {Aldrich}, C.~H., {Forslund}, D.~W., \& {Quest}, K.~B. 1986,
  \prl, 56, 1059

\bibitem[{{Treumann} {et~al.}(2004){Treumann}, {Jaroschek}, \&
  {Scholer}}]{treumann04}
{Treumann}, R.~A., {Jaroschek}, C.~H., \& {Scholer}, M. 2004, Physics of
  Plasmas, 11, 1317

\bibitem[{{Yoon}(2011)}]{yoon11}
{Yoon}, P.~H. 2011, Physics of Plasmas, 18, 122303

\bibitem[{{Yoon}(2012)}]{yoon12}
{Yoon}, P.~H. 2012, Physics of Plasmas, 19, 012304

\bibitem[{{Zaheer} \& {Yoon}(2013)}]{zaheer13}
{Zaheer}, S. \& {Yoon}, P.~H. 2013, \apj, 775, 108

\bibitem[{{Zank} {et~al.}(2010){Zank}, {Heerikhuisen}, {Pogorelov}, {Burrows},
  \& {McComas}}]{zank10}
{Zank}, G.~P., {Heerikhuisen}, J., {Pogorelov}, N.~V., {Burrows}, R., \&
  {McComas}, D. 2010, \apj, 708, 1092

\end{thebibliography}
\end{document}